\documentclass[prl,twocolumn,superscript address]{revtex4}

\usepackage{epic,eepic,amsmath,latexsym,amssymb,verbatim,bbm, graphicx,epic,eepic,epsfig, color}

\usepackage{pstricks}
\newcommand{\nc}{\newcommand}
\nc{\rnc}{\renewcommand}
\nc{\beq}{\begin{equation}}
\nc{\eeq}{{\end{equation}}}
\nc{\beqa}{\begin{eqnarray}}
\nc{\eeqa}{\end{eqnarray}}
\nc{\lbar}[1]{\overline{#1}}
\nc{\bra}[1]{\langle#1|}
\nc{\ket}[1]{|#1\rangle}
\nc{\ketbra}[2]{|#1\rangle\!\langle#2|}
\nc{\braket}[2]{\langle#1|#2\rangle}
\nc{\proj}[1]{| #1\rangle\!\langle #1 |}
\nc{\avg}[1]{\langle#1\rangle}
\rnc{\max}{\operatorname{max}}
\nc{\Rank}{\operatorname{Rank}}
\nc{\smfrac}[2]{\mbox{$\frac{#1}{#2}$}}
\nc{\Tr}{\operatorname{Tr}}
\nc{\id}{\operatorname{id}}
\nc{\1}{\openone}
\nc{\ox}{\otimes}
\nc{\dg}{\dagger}
\nc{\dn}{\downarrow}
\nc{\cA}{{\cal A}}
\nc{\cB}{{\cal B}}
\nc{\cC}{{\cal C}}
\nc{\cD}{{\cal D}}
\nc{\cE}{{\cal E}}
\nc{\cF}{{\cal F}}
\nc{\cG}{{\cal G}}
\nc{\cH}{{\cal H}}
\nc{\cI}{{\cal I}}
\nc{\cJ}{{\cal J}}
\nc{\cK}{{\cal K}}
\nc{\cL}{{\cal L}}
\nc{\cM}{{\cal M}}
\nc{\cN}{{\cal N}}
\nc{\cO}{{\cal O}}
\nc{\cP}{{\cal P}}
\nc{\cR}{{\cal R}}
\nc{\cS}{{\cal S}}
\nc{\cT}{{\cal T}}
\nc{\cX}{{\cal X}}
\nc{\cZ}{{\cal Z}}
\nc{\supp}{{\operatorname{supp}}}
\nc{\var}{\operatorname{var}}
\nc{\rar}{\rightarrow}
\nc{\lrar}{\longrightarrow}
\nc{\polylog}{\operatorname{polylog}}

\def\d{\delta}
\def\e{\epsilon}
\def\ve{\varepsilon}

\nc{\RR}{{{\mathbb R}}}
\nc{\CC}{{{\mathbb C}}}
\nc{\FF}{{{\mathbb F}}}
\nc{\NN}{{{\mathbb N}}}
\nc{\ZZ}{{{\mathbb Z}}}
\nc{\PP}{{{\mathbb P}}}
\nc{\QQ}{{{\mathbb Q}}}
\nc{\UU}{{{\mathbb U}}}
\nc{\EE}{{{\mathbb E}}}

\nc{\newOne}{{{\mathbb{I}}}}

\nc{\be}{\begin{equation}}
\nc{\ee}{{\end{equation}}}
\nc{\bea}{\begin{eqnarray}}
\nc{\eea}{\end{eqnarray}}
\nc{\<}{\langle}
\rnc{\>}{\rangle}
\nc{\Hom}[2]{\mbox{Hom}(\CC^{#1},\CC^{#2})}
\nc{\rU}{\mbox{U}}

\begin{document}
\psset{unit=0.25mm}
\author{Joseph M.~Renes}
\affiliation{Institut f\"ur Angewandte Physik, Technische Universit\"at Darmstadt, D-64289 Darmstadt, Germany}
\author{Graeme Smith}
\affiliation{Institute for Quantum Information, Caltech 107--81,
    Pasadena, CA 91125, USA}

\title{Noisy Processing and the Distillation of Private States}

\begin{abstract}
We provide a simple security proof for prepare \& measure quantum key distribution protocols
employing noisy processing and one-way postprocessing of the key.  This is achieved by showing that the
security of such a protocol is equivalent to that of an associated key distribution protocol
in which, instead of the usual maximally-entangled states, a more general {\em private state} is distilled.  
Besides a more general target state, the usual entanglement distillation tools are employed (in particular, Calderbank-Shor-Steane (CSS)-like codes), 
with the crucial difference that noisy processing allows some phase errors to be left uncorrected without compromising the privacy of the key.

\end{abstract}

\date{\today}
\maketitle

Entanglement has been the cornerstone of many 
quantum key distribution (QKD) security proofs to date: A 
prepare \& measure protocol by which Alice and Bob generate a secret key can be
shown to be secure exactly when an associated entanglement distillation protocol 
succeeds in producing a high fidelity maximally-entangled state.
Secrecy of the key then follows since maximal entanglement can only be shared between two 
parties~\cite{LoChau99,ShorPreskill00,Lo01,GP01,TKI03,GL03,BTBLR05,RG05}. 
The resulting proofs are 
intuitive and allow QKD designers to
incorporate current methods of quantum error correction and entanglement distillation.

Renner, Gisin, and Kraus adopt a more information-theoretic approach 
to QKD security with the surprising result that secure key can be established at noise levels beyond what 
seems possible in the entanglement-based picture \cite{KGR}.  By including a step in which Alice adds noise to her sifted key
the overall key rate can actually increase. The additional 
noise damages the correlations held by Alice and Bob but the key observation
is that this noise may damage Eve's correlations even more.  
While the best known upper bounds for one-way distillable entanglement
do not rule out the possiblity of distilling EPR pairs for these noise levels,  it is puzzling that this processing can 
generate key at rates well in excess of the best known entanglement distillation rates~\cite{SmithSmo0506}.  Thus, it has been unclear whether an entanglement-based security proof is possible for these protocols.

We find a resolution in the observation of \cite{HHHO05}
that maximally-entangled states are not strictly necessary for generating secret
keys.  Instead, states leading to secret keys belong
to the class of {\em private states}.  These are composed of completely correlated 
systems $A$ and $B$ containing a uniformly distributed key, along with
{\em shield} systems $A'$ and $B'$.  More precisely, $\gamma$ is called 
a $d$-dimensional private state (or {\em pdit}) if there are unitaries
$U^{(j)}$ and a {\em twisting operator} of the form 
$U_{\rm twist}=\sum_{j}\ket{jj}\bra{jj}_{AB}\otimes U^{(j)}_{A'B'}$,
such that $\gamma =U_{\rm twist}\left(\proj{\Phi_d}_{AB}\otimes \rho_{A'B'}\right) U_{\rm twist}^\dagger$
for some $\rho_{A'B'}$, where $\ket{\Phi_d} =\sum_{i=1}^d \ket{ii}/\sqrt{d}$.
The twisting operator ensures that, while Alice and Bob may not share a maximally entangled state, 
Eve's reduced state is independent of the key. This definition
recalls an earlier result \cite{AB02} that the secrecy of 
key created from entangled systems is not diminished by phase noise in the 
devices performing the entanglement distillation. 

\begin{figure}[htbp]
\begin{center}
\begin{pspicture}(-30,-127)(280,5)
\rput[l](260,0){\mbox{$A^\prime$}}
\rput[l](260,-20){\mbox{$A$}}
\rput[l](260,-60){\mbox{$B$}}
\rput[l](260,-100){\mbox{$E_1$}}
\rput[l](260,-120){\mbox{$E_2$}}
\rput[r](-5,0){\mbox{$\ket{\varphi_q}$}}
\rput[r](-5,-25){\mbox{$\ket{\Phi}$}}
\rput[r](-5,-110){\mbox{$\ket{\eta_{uv}}$}}
\rput(110,-60){\mbox{$Z$}}
\psline[linewidth=.5pt](0,0)(250,0)
\psline[linewidth=.5pt](0,-20)(250,-20)
\psline[linewidth=.5pt](0,-30)(20,-30)(40,-60)(100,-60)
\psline[linewidth=.5pt](120,-60)(250,-60)
\psline[linewidth=.5pt](0,-100)(250,-100)
\psline[linewidth=.5pt](0,-120)(250,-120)
\pscircle[linewidth=.5pt](110,-20){5.5}
\pspolygon[linewidth=.5pt](100,-50)(120,-50)(120,-70)(100,-70)
\pscircle[linewidth=.5pt](145,-60){5.5}
\psline[linewidth=.5pt](110,0)(110,-25.5)
\psline[linewidth=.5pt](110,-70)(110,-100)
\psline[linewidth=.5pt](145,-54.5)(145,-120)
\qdisk(145,-120){1.25pt}
\qdisk(110,-100){1.25pt}
\qdisk(110,0){1.25pt}
\end{pspicture}
\end{center}
\vspace{-.25cm}
\caption{\label{figure:twisted}
The effective state held by Alice, Bob, and Eve after noisy processing, where 
$\ket{\varphi_q} = \sqrt{1{-}q}\ket{0}{+}\sqrt{q}\ket{1}$ ,
$\ket{\eta_{uv}} = \sum_{u,v}\sqrt{p_{uv}}\ket{uv}$
and $A^\prime$ is the purification of the noise Alice adds.
CSS-like error correction on the $AB$ system is equivalent to 
classical error correction and privacy amplification on the key in the
prepare \& measure protocol, and securely provides key exactly 
when it maps many copies of the above state to a high-fidelity private state for all $p_{uv}$ consistent with the estimated parameters.  The shield consists of the $A^\prime$ systems together with the CSS code's syndrome bits held by Bob. }
\end{figure}
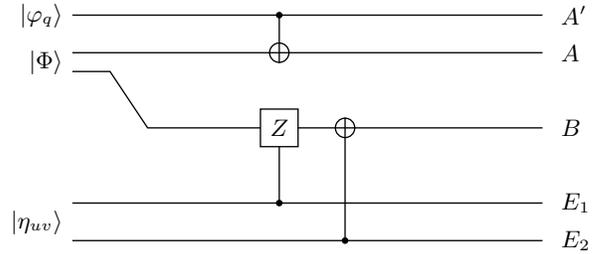

In \cite{HLLO05} it was shown that a large number of low fidelity 
copies of a private state
can sometimes
be distilled to a high fidelity private state with the same shield but smaller key system.
However, it is not clear what class of
QKD protocols can be coherently recast in the form considered by \cite{HLLO05}.
As we will see, the coherent version of \cite{KGR}'s protocol is quite different from those of \cite{HLLO05}---the initial
adversarially distributed state will be noisy EPR pairs (with no shield), and the shield of the final private state arises due to Alice and Bob's noisy processing.

In the following, we show that a prepare \& measure QKD scheme with noisy processing
and one-way postprocessing is secure exactly when an 
associated pdit distillation protocol has high
fidelity.  This requires only minor modifications of the standard entanglement distillation argument.
Indeed, in the coherent description of the noisy processing protocol the auxiliary system purifying the noise introduced by Alice will function as a shield, 
and the sifted key will become noisy EPR pairs in the key system of a noisy pdit.
The error correction and privacy amplification required in the classical processing
maps to a CSS-like  quantum code on 
the key system in the coherent protocol in the same way as found in \cite{ShorPreskill00}.
If the CSS code performs a suitable amount of bit and phase
error correction on the key system, Alice and Bob will be left with a high fidelity private state. 
The crucial difference from previous entanglement-based security proofs is that 
\emph{Alice and Bob need not correct every phase error to guarantee security}, and this savings will often more than 
compensate for the associated increase in the number of bit errors they must correct.
In fact, we can establish key at bit error rates up to $12.4\%$ for the Bennett-Brassard-84 (BB84) protocol \cite{BB84} and $14.1\%$ for the 
six-state protocol, matching the rates of \cite{KGR} and surpassing all previous thresholds from entanglement-based proofs.

\emph{Private State Distillation.}---We begin with a coherent reformulation of the
BB84 and six-state protocols~\cite{LoChau99,ShorPreskill00};
other protocols can be handled in a similar manner \cite{RG05}. 
In both cases, Alice first prepares the state $\ket{\Phi}_{AB}$ 
and sends the $B$ system to Bob. In BB84 (six-state), each party then randomly measures
in the $X$ or $Z$ basis ($X$, $Y$, or $Z$) and by public discussion 
they sift out those outcomes corresponding to the same basis choice.
This is equivalent to Alice (Bob) sending a random 
bit in (measuring in) one of the bases at random,
since the statistics of measurements as well 
as an eavesdropper Eve's dependence on their outcomes 
are identical in both cases.  Alice and Bob then 
publicly compare a small fraction of the sifted key
to estimate the noise level of the channel.

If the noise level is zero, the resulting length-$n$ sifted key can be described coherently as
$\ket{\Phi}^{\otimes n}$.  Otherwise, the most general noisy channels we
need to consider are Pauli channels, since all subsequent 
operations will commute with a (hypothetical) measurement in the Bell-basis
which digitizes the actual noise into this form~\cite{LoChau99,GL03}. 
The only difference here to the original classicization argument of Lo and Chau
is that Alice flips some key bits, which also commutes with the Bell-state 
measurement. Attributing the noise to Eve, 
the key state is
\begin{equation}
\label{keystate1}
\sum_{\mathbf{u, v}} \sqrt{p_\mathbf{u v}} \, 
(\newOne_A\ox X^{\mathbf u}_B Z^{\mathbf v}_B)
\ket{\Phi}_{AB}^{\ox n}\ket{\mathbf u}_{E_1}\ket{\mathbf v}_{E_2},
\end{equation}
where $p_\mathbf{u v}$ is the probability of error pattern 
$X^\mathbf{u}Z^\mathbf{v}$ described by length-$n$ bit strings $\mathbf{u}$ and 
$\mathbf{v}$. Furthermore, if Alice and Bob randomly permute their $n$ systems, it is 
sufficient to consider noise that is independent and 
identically-distributed (i.i.d) for each transmitted qubit,
given by rate $p_{uv}$. This follows from a slight variant of Lemma 3 of \cite{GL03} 
(see also \cite{ShorPreskill00,GP01}), the particulars of which we take up after the 
detailed analysis of the next section. 

Alice and Bob now distill the key
by performing bit error-correction and privacy amplification (phase error-correction).
Before this, Alice adds i.i.d.~noise to A, randomly applying $X$ at rate $q$. 
This is described coherently as using an auxiliary system $A'$ in the state 
$\ket{\varphi}_{A'}=\sqrt{1-q}\ket{0}{+}\sqrt{q}\ket{1}$
as the control system in a CNOT gate, yielding the state
\begin{equation}
\label{keystate2}
\sum_{\mathbf{u, v,f}} \sqrt{p_\mathbf{u v}q_\mathbf{f}} \, 
\ket{\mathbf{f}}_{A'}(X^\mathbf{f}_A\ox X^{\mathbf u}_B Z^{\mathbf v}_B)
\ket{\Phi}_{AB}^{\ox n}\ket{\mathbf u}_{E_1}\ket{\mathbf v}_{E_2},
\end{equation}
where $q_\mathbf{f}=q^{|\mathbf{f}|}(1-q)^{n-|\mathbf{f}|}$ 
for length-$n$ bit string $\mathbf{f}$ and $|\mathbf{f}|$ its Hamming weight. 
We can also think of Alice's error operator acting on Bob's system, since $X\otimes XZ$
and $\newOne\otimes XZX$ have the same effect on $\ket{\Phi}$.

Now Alice and Bob perform bit error-correction using a linear error correcting code.
This step is the same as the usual analysis, since all bit errors must be 
corrected in the final key.  The bit error rate is $\tilde{p} = p_x(1-q) + q(1-p_x)$ for $p_x=\sum_v p_{1,v}$, 
so Alice and Bob must measure $n H_2(\tilde{p})$ parity syndromes, where $H_2$ 
is the binary Shannon entropy, in order to identify the
error pattern with high probability. 
To simplify the resulting expressions, we use the method of 
decoupling error correction and privacy amplification~\cite{Lo03},
itself based on the breeding entanglement distillation protocol~\cite{BBPSSW96},
whereby syndromes are collected in auxiliary entangled pairs.

Alice collects the bit parities in her halves of the ancilla states,
measures them, and sends the result to Bob. 
Bob then coherently corrects system $B$
and records the error in an ancilla system $B'$, producing
\begin{equation}
\label{keystate3}
\sum_{\mathbf{u, v,f}} \sqrt{p_\mathbf{u v}q_\mathbf{f}} \, 
Z^\mathbf{v}_{A'}\ket{\mathbf{f}}_{A'}\ket{\mathbf{u}+\mathbf{f}}_{B'} Z^{\mathbf v}_B
\ket{\Phi}_{AB}^{\ox n}\ket{\mathbf u}_{E_1}\ket{\mathbf v}_{E_2},
\end{equation}
where $Z^{\mathbf v}_{A'}$ comes from interchanging 
$X^{\mathbf f}_{B}$ and $Z^{\mathbf v}_{B}$.

In the classical description of the protocol, this step requires Alice to encrypt her measurement outcomes
with a one-time pad, preventing information leakage to Eve.
This encryption requires a key, which in the coherent description is a private state,
meaning Alice and Bob generally collect the parity syndromes 
in the key subsystems of private states, not in maximally-entangled
states as we have used. However, there is no loss of generality in using maximally-entangled
states in the formalism, since using private states raises no additional complications~\cite{HLLO05,NoteOnAnc}.

At this stage, the normal entanglement-based proof would proceed to correct all phase
errors.  This would not give the key rates of~\cite{KGR}
as the extra noise would just reduce 
the rates from those of \cite{ShorPreskill00}.  Instead, we come to the  
main observation of this paper: not all phase errors must be corrected. After correcting
enough, the resulting state will be close to a private state.

Examining Alice and Bob's state makes 
clear how this comes about. Tracing out Eve's systems, they hold 
\begin{equation}
\label{state3}
\rho=C_{A'B'}\left(\sum_{\mathbf{u},\mathbf{v}}p_{\mathbf{u} \mathbf{v}}[\mathbf{u}]_{B'}
[\varphi^\mathbf{v}]_{A'} Z^\mathbf{v}_B[\Phi]_{AB}^{\ox n}Z^\mathbf{v}_B\right)C_{A'B'}^\dagger,
\end{equation}
where $[\theta]=\ketbra{\theta}{\theta}$, $\ket{\varphi^\mathbf{v}}=Z^\mathbf{v}\ket{\varphi}^{\otimes n}$, and we have used a CNOT
$C_{A'B'}$ to write $\ket{\mathbf{f}}_{A'}\ket{\mathbf{u}{+}\mathbf{f}}_{B'}$ as 
$C_{A'B'}\ket{\mathbf{f}}_{A'}\ket{\mathbf{u}}_{B'}$.

By performing phase error correction at a reduced rate, the pattern of phase errors will not be 
uniquely identified, but only narrowed to a set 
$\mathcal{V}_\mathbf{s}$ indexed by the syndrome $\mathbf{s}$: 
$\mathcal{V}_\mathbf{s}=\{\mathbf{v}\,|\,{\rm syndrome}(\mathbf{v})=\mathbf{s}\}$.
The key point is that if the vectors $\ket{\varphi^\mathbf{v}}$ for $\mathbf{v}\in \mathcal{V}_\mathbf{s}$
were mutually orthogonal, we could define the unitary 
$D_{A'B}=\sum_\mathbf{v\in \mathcal{V}_s}[\varphi^\mathbf{v}]_{A'}\otimes Z^\mathbf{v}_B$
and use $U_{BA'B'}=D_{A'B}C_{A'B'}$ to untwist:
\begin{eqnarray}
\rho'&=&U_{BA'B'}\rho U_{BA'B'}^\dg\nonumber\\
&=&[\Phi]^{\otimes n}_{AB}\otimes\Big(\sum_\mathbf{u}p_\mathbf{u}[\mathbf{u}]_{B'}
\sum_{\mathbf{v}\in\mathcal{V}_s}p_{\mathbf{v}|\mathbf{u}}\,
[\varphi^\mathbf{v}]_{A'}\Big).
\end{eqnarray}
Since $D$ is a controlled-$Z$ gate, either system can be thought of as the control, 
so $D_{A'B}=\sum_\mathbf{j} U^{(\mathbf{j})}_{A'}\otimes [\mathbf{j}]_B$ 
for some unitaries $U^{(j)}$.  $U_{BA'B'}$ is a twisting operation, so that Alice and Bob would share a
private state. Keys derived from this state would be secret.

\emph{Detailed Analysis.}---To establish the secrecy of 
keys generated from $\rho$, recall the univerally-composable
definition of security from \cite{RK05,BHLMO05}. A key $K$ is called $\epsilon$-secure if the 
state $\rho_{KE}$ of the key and eavesdropper satisfies $||\rho_{KE}-\kappa\otimes \rho_E||_1\leq 2 \epsilon$,
where $\kappa$ is a uniform mixture of all key values, shared by Alice and Bob. 
The latter state is a perfect key and this definition ensures that
$\rho_{KE}$ can safely be used for any further cryptographic purpose. 
 
In the present context, the key is created by measuring systems $A$ and $B$ of $\rho$ in the $Z$ basis.
As the untwisting operation is unitary and commutes with the 
measurement, whether it is performed before the measurement or after 
does not affect the key's security.  When performing the untwisting operation on the unmeasured state 
results in a maximally-entangled state on $AB$, the key generated will be perfectly secure. 
Similarly, if there is an untwisting operation 
mapping $AB$ to within $2\epsilon$ of a maximally-entangled state,
the key is $\epsilon$-secure \cite{HHHO05}.

For simplicity we consider independent amplitude and phase errors, with the case of correlated 
$\mathbf{u}$ and $\mathbf{v}$ following along similar lines.  
To construct an untwisting operation, it suffices to find a 
rank-one POVM with elements $E_\mathbf{v}$ that can distinguish
the $\ket{\varphi^\mathbf{v}}$ with average error $P_{\rm e}$ no larger than $\epsilon^2/2$: 
$
P_{\rm e}=\langle P_{\rm e}^\mathbf{v}\rangle=\sum_{\mathbf{v},\mathbf{v}'\neq \mathbf{v}} 
p_\mathbf{v} \bra{\varphi^\mathbf{v}} E_{\mathbf{v}'}
\ket{\varphi^\mathbf{v}} \leq \epsilon^2/2,
$
where $P_{\rm e}^\mathbf{v}$ 
is probability of decoding input state $\ket{\varphi^\mathbf{v}}$ incorrectly.  
This problem was considered by 
\cite{HJSWW96} in the context of transmitting classical information over a quantum channel.
Letting $\sigma=(1-p_z)\ketbra{\varphi}{\varphi}+p_z Z\ketbra{\varphi}{\varphi}Z$, $p_z = \sum_{u}p_{u1}$, and $S(\sigma)$
be the entropy of $\sigma$,
their results imply that with probability $1-\epsilon^2/2$, the 
elements of a randomly-chosen subset $\mathcal{V}_{\mathbf s}\subset \mathcal{V}$ of size $2^{n(S(\sigma)-\delta)}$ can 
be distinguished by the pretty-good measurement (PGM) with average error probability $\e^2/2$,
where $\epsilon$ decreases exponentially with $n$ for arbitrarily small positive $\delta$. 

The PGM has rank-one elements by construction~\cite{HW94}, so we have
$E_\mathbf{v}=\ketbra{\widetilde{\theta}^\mathbf{v}}{\widetilde{\theta}^\mathbf{v}}$ 
for unnormalized $\ket{\widetilde{\theta}^\mathbf{v}}$. Then we can append another
auxiliary system $A''$ and consider the Neumark extension
consisting of orthonormal states $\ket{\theta^\mathbf{v}}_{A'A''}$ in the joint Hilbert space $A'A''$
such that $_{A'A''}\braket{\theta^\mathbf{v}}{\varphi^\mathbf{v'}}_{A'}\ket{0}_{A''}=
\,_{A'}\braket{\widetilde{\theta}^\mathbf{v}}{\varphi^\mathbf{v}}_{A'}$~\cite{NC00}.
With this, we can finally construct the untwisting operator 
$U=(\sum_{\mathbf{v}}[\theta^\mathbf{v}]_{A'A''}\otimes Z^\mathbf{v}_B)C_{A'B'}^\dagger$.

Letting $\widetilde{\rho}=\proj{0}\ox \rho$, the fidelity of $U\widetilde{\rho}\, U^\dagger$ with 
$\rho'=[\Phi]^{\otimes n}_{AB}\otimes \sum_{\mathbf{u},\mathbf{v}} p_{\mathbf{u},\mathbf{v}}
[\theta^\mathbf{v}]_{A'A''}\otimes [\mathbf{u}]_{B'}$ is given by
$
F(U\widetilde{\rho}\,U^\dagger,\rho')=\sum_{\mathbf{u},\mathbf{v}} p_{\mathbf{u}\mathbf{v}}
\,|\braket{\varphi^\mathbf{v}}{\widetilde{\theta}^\mathbf{v}}|
=\langle \sqrt{P_{\rm s}^\mathbf{v}}\rangle,
$
where $P_{\rm s}^\mathbf{v}$ is the conditional probability of successful transmission of $\mathbf{v}$.
Since $\langle \sqrt{P_{\rm s}^\mathbf{v}}\rangle\geq \langle P_{\rm s}^\mathbf{v}\rangle=1-P_{\rm e}\geq 1-\epsilon^2/2$,
using the relation between trace norm and fidelity \cite{FvG99}, we find 
$||U\widetilde{\rho}\, U^\dagger-\rho'||_1\leq  2\sqrt{1-F^2}\leq  2\e$, proving $\e$-security.

  A subtlety arises in the use of the Neumark extension in that our untwisting operation
consists of controlled isometries rather than unitaries.  However, the privacy of the key is uncompromised: while Eve may have knowledge of the shield system,
as long as Alice and Bob hold the key and shield, the fact that they could be untwisted implies that
Eve is ignorant of the key.

Above, we took $\mathbf u$ and $\mathbf v$ to be independent.  If they are not, 
randomly choosing sets ${\cal V}_{\mathbf s}$ of size $2^{n(S(\sigma |u)-\d)}$, 
where $S(\sigma|u)$ is the conditional entropy of $\sigma$
given $u$, leads to an exponentially small average probability of decoding error for the PGM, 
and the rest of the argument remains unchanged\cite{Lo01}.
Putting this all together, by using a random code Alice and Bob can select a subset $\mathcal{V}_{s}$
of size $\approx 2^{nS(\sigma|u)}$. With probability exponentially close to one, 
the untwisting operation can be constructed from the pretty-good measurement,
ensuring the key is $\e$-secure.

Finally, we must consider the effects of non-i.i.d.~noise, e.g.~arising from a coherent 
eavesdropping attack. By random sampling Alice and Bob 
obtain an estimate $f^{\rm est}_{u,v}$ of
the fraction, or \emph{type}, of Pauli errors $X^uZ^v$. 
Since the raw key bits are permutation-invariant, 
$|f^{\rm est}_{u,v}{-}f^{\rm true}_{u,v}|\leq \ve$ with
probability exponentially close (in $n$) to unity~\cite{KR05}. 
This allows us to prove that the above procedure is secure for \emph{any} input state
yielding estimate $f^{\rm est}_{u,v}$, not just those subjected to i.i.d.~noise. 
First decompose the squared fidelity for an i.i.d.~input state with error 
rate $p_{u,v}{=}f^{\rm est}_{u,v}$ into a sum over 
possible types $f$: 
$F^2=\sum_{f}{\rm prob}(f|p{=}f^{\rm est})F^2_{f}$, where
prob$(f|p{=}f^{\rm est})$ is the probability of type $f$ in the i.i.d.~distribution, $F_f$ is the
fidelity our protocol produces on a uniform distribution over errors of type $f$, 
and we have suppressed the $u,v$ indices.
Since there are only polynomially many types, all those with 
nonnegligible probability must have polynomially large probability and thus 
corresponding fidelities $F_f$ which are
exponentially close to one. Since types within $\ve$ of the rate $p$ are among
the probable types~\cite{CK81}, this guarantees that the above procedure 
produces high-fidelity entangled
output states (or securely aborts) for any input state yielding $f^{\rm est}$. 

\emph{Achievable Key Rates.}---What key generation 
rates can be achieved by the protocols considered above?
The bit-error correction step consumes $n H_2(\tilde{p})$ 
previously-established secret key bits, but in so doing produces 
$n$ bit-error-free bits. The phase error correction must 
reduce the number of phase errors from $2^{nH(v|u)}$ 
to $2^{nS(\sigma|u)}$ (which can be accomplished by a 
random phase code with $n(H(v|u)-S(\sigma|u))$ syndrome bits) in
order to ensure that Alice and Bob could untwist the state, 
so we find an overall rate of $1 - H_2(\tilde{p}) - (H(v|u)-S(\sigma|u))$, or
\begin{equation}
R=1-H_2(\tilde{p})-\sum_u p_u \left(H_2(p_{1|u})-H_2(\lambda^+_u)\right),
\end{equation}
where $\lambda^+_u = \frac{1}{2}(1+\sqrt{1-16q(1-q)p_{1|u}(1-p_{1|u})})$
is the larger eigenvalue of 
$\sigma_u = (1-p_{1|u})\proj{\varphi} + p_{1|u}Z\proj{\varphi}Z$.

In the BB84 protocol, bit and phase errors are equal but uncorrelated, meaning $p_{1|u}=p_z = p_x = p_{1|v}$, from which we find an
error threshold of $12.4\%$ by letting $q\rightarrow 1/2$. In the six-state protocol 
all Pauli errors occur at the same rate, giving
a threshold error rate of $14.1\%$. 

\emph{Discussion}---We have shown that one-way key distribution 
protocols employing noisy processing can be seen
as distillation protocols for private states where the purification
of the added noise functions as part of the shield and the error correction 
and privacy amplification steps map to a CSS code in the usual way.
This extends the entanglement distillation paradigm 
initiated in \cite{LoChau99,ShorPreskill00}, providing a cleaner and less 
technical security proof for the protocols of \cite{KGR}.  Further, by formulating the protocol in this way, we gain insight into the mechanism by which addition of noise improves key rates, namely by 
deflecting Eve's correlations with Alice and Bob to the shield and away from the key.

In the security proof of the six-state protocol~\cite{Lo01}, building on the work of \cite{DSS98}, 
Lo showed that a degenerate error-correcting code could be used to improve the 
threshold error rate from $12.6\%$ to $12.7\%$.  Further progress in this direction 
can be found in \cite{SRS06}, where we report on the combination of that method with the noisy processing studied here, showing that the threshold error rate of BB84 can be increased from $12.4\%$ to $12.9\%$.  We believe our findings will point towards new methods of key distillation and analagous methods of private state distillation, furthering the fruitful exchange between privacy amplification 
and entanglement distillation.

{\em Acknowledgements}---We thank D. Leung, G. O. Myhr, 
G. Nikolopoulos, R. Renner, and B. Toner for helpful discussions. 
This work was initiated at the University of Queensland, 
and we are grateful to M. Nielsen for his hospitality.
JMR supported by the Alexander von Humboldt foundation 
and the European IST project SECOQC, and GS by
NSF grant PHY-0456720 and Canada's NSERC.


\begin{thebibliography}{30}
\bibitem{NoteOnAnc}
When using the key part of a private state to store the
  result of measuring a
  bit parity of a collection of noisy EPR pairs, the control from the key system to the shield
  is transfered to the noisy EPR pairs. This implies that the noisy EPR pairs
  can be corrected to a private state exactly when using a perfect EPR to
  measure the parity would have made this possible.
\end{thebibliography}
\end{document}